\documentclass[fleqn,usenatbib]{mnras}

\usepackage{newtxtext,newtxmath}

\usepackage[T1]{fontenc}
\usepackage{ae,aecompl}

\usepackage{graphicx}	
\usepackage{amsmath}	
\usepackage{amssymb}	
\usepackage{pdflscape}

\usepackage[normalem]{ulem} 

\newcommand{\aro}{{$\alpha_{\rm ro}$~}}
\newcommand{\aox}{{$\alpha_{\rm ox}$~}}
\newcommand{\nupeak}{$\nu_{peak}$~}
\newcommand{\ee}{\end{equation}}
\newcommand{\be}{\begin{equation}}
\newcommand{\ergj}{{erg~cm$^{-2}$s$^{-1}$Jy$^{-1}$~}}
\newcommand{\fxfr}{$f_{\rm x}/f_{\rm r}$~}

\title[Deep \textit{Swift}/SDSS blazars sample]{A new sample of X-ray selected \textit{Swift}/SDSS faint blazars and blazar candidates}

\author[S. Turriziani et al.]{
Sara Turriziani,$^{1}$\thanks{E-mail: sara.turriziani@riken.jp}
B. Fraga,$^{2}$
and P. Giommi$^{3,4,5}$
\\
$^{1}$Computational Astrophysics Laboratory - RIKEN, 2-1 Hirosawa, Wako, Saitama 351-0198, Japan\\
$^{2}$Centro Brasileiro de Pesquisas F\'{i}sicas, rua Dr. Xavier Sigaud 150, 22290-180, Rio de Janeiro, Brazil\\
$^{3}$Italian Space Agency, ASI, via del Politecnico s.n.c., 00133 Roma, Italy \\
$^{4}$Institute for Advanced Studies,  Technische Universit\"at M\"unchen, Lichtenbergstrasse 2a, D-85748 Garching bei M\"unchen, Germany\\
$^{5}$ICRANet, P.zza della Repubblica 10, 65122, Pescara, Italy
}

\date{Accepted XXX. Received YYY; in original form ZZZ}

\pubyear{2015}

\begin{document}
\label{firstpage}
\pagerange{\pageref{firstpage}--\pageref{lastpage}}
\maketitle

\begin{abstract}
Our present knowledge of the properties of blazars mostly comes from small samples of bright objects, especially regarding studies on their cosmological evolution. 
Statistically well defined and completely identified samples of faint blazars are very difficult to obtain.
We present a new X-ray selected sample of 62 blazars and blazar candidates reaching deep X-ray fluxes. We relied on the availability of large catalogs of astronomical objects combined with on-line services offering simple access to finding charts and magnitude estimates. We built the sample cross-matching X-ray sources in the \textit{Swift} Serendipitous Survey in deep XRT GRB Fields catalog with data from deep radio and optical surveys. 
Our sample can probe populations of sources 10 times weaker in the X-ray flux with respect previous studies, thus allowing for a more detailed comparison between data and simulated counts.
We use the sample to calculate the radio and X-ray LogN-LogS of blazars down to fluxes at least one order of magnitude fainter than previous studies. We show that, considering that our sample may be somewhat contaminated by sources other than blazars, we are in agreement with previous observational and theoretical estimations.
\end{abstract}

\begin{keywords}
Galaxies: active --  Galaxies: BL Lacertae objects: -- X-ray : galaxies -- Catalogs -- methods: statistical -- Radiation mechanisms: non-thermal 
\end{keywords}




\section{Introduction}

Surveys often played a crucial role in achieving significant progress in astronomical research. This is because they provide the observational data that hold the statistical 
information needed to characterize the underlying source populations. Besides real X-ray surveys such as RASS BSC and FSC \citep{rosatbright,rosatfaint} and the most recent 2RXS \citep{2rxs}, a fundamental role is played by serendipitous surveys. In fact, serendipitous X-ray surveys exploit the relatively wide field of view of typical X-ray imaging instrumentation by searching for sources that happen to be located nearby the target of pointed observations. Such surveys are quite common and have been carried out with most X-ray satellites since the Einstein observatory was launched. The resulting serendipitous source catalogs - e.g. EMSS \citep{emss}, WGACAT \citep{wgacat} -
served as the basis for numerous studies and gave a significant contribution to understand the nature of various Galactic and extragalactic source populations.

Mining survey data can be therefore crucial to advance our current knowledge of blazars, a powerful and rare class of Active Galactic Nuclei (AGN). Blazars are characterized by a strong and highly variable, non-thermal emission from radio wavelengths up to TeV energies, showing a typical double-peaked spectral energy distributions (SEDs). As initially suggested by \citet{blandford}, the peculiarities of this emission (e.g.  flat radio spectral index, superluminal motion, high brightness temperatures) can be explained by relativistic amplification, since we are observing a collimated jet of energetic particles pointing toward our direction.

There are different way to classify blazars. For example, on the basis of the appearance of their optical spectrum, they are conventionally divided into two main subclasses: Flat Spectrum Radio Quasars (FSRQs), and BL Lac Objects (BL Lacs). FSRQs show in fact prominent emission lines such as other quasars, whereas BL Lacs have no or really weak emission lines.

An alternative and complementary classification scheme uses the position of the first peak of the SED, attributed to synchrotron emission, to distinguish between: i) Low Synchrotron Peaked blazars (LSPs), when the synchrotron peak is in the IR/far-IR band (\nupeak $< 10^{14} Hz$); ii) High Synchrotron Peaked blazars (HSPs) when this peak moves to UV or higher energies (\nupeak $> 10^{15} Hz$); and iii) Intermediate Synchrotron Peaked blazars (ISPs) in the intermediate cases \citep{P95, abdosed10}.

\par Although blazars are a small fraction of the overall AGN population, they contribute significantly to the cosmic extragalactic backgrounds in those frequency bands where the accretion mechanism does not produce radiation \citep[e.g.][]{giommicmb}.
\par Our present knowledge of blazars comes from relatively small samples, especially regarding studies on their cosmological evolution. So far, many efforts have been made to define larger blazar samples in order to better constrain the peculiar nature of these sources, their multi-frequency properties, their statistics, evolution with cosmic time and their contribution to background radiations, specially in the microwave and $\gamma$-ray bands (e.g. Sedentary,  \citet{sedentary}; DXRBS, \citet{deepxrayradio}; ROXA, \citet{roxa}; WIBRaLS, \citet{WGS2014}; 1WHSP,  \citet{1whsp}; \citet{addrecent1}; \citet{addrecent2}). In order to enhance our knowledge of blazars, it is necessary to have complete samples down to very faint fluxes.
\par In this context, the Neil Gehrels \textit{Swift} Observatory \citep[hereafter, \textit{Swift};][]{Swift1} provides unique capabilities. Although it was designed to discover Gamma-Ray Bursts (GRBs), the findings made by its telescope are transcending the science of GRBs and have a broad impact in astronomical research, with many scientists using \textit{Swift} data for their works \citep{madridmacchetto, Savaglio2013}. As of today, \textit{Swift} discovered over 1,000 GRBs, a large fraction of which have been followed for several days. This makes the GRB fields of \textit{Swift} a good dataset to look for serendipitous faint X-ray sources; moreover, any catalog built with these pointings would be unbiased, since GRBs are thought to explode randomly across the sky and blazars are totally unrelated to these sources (while the same may not be  true for other types of extragalactic targets). 
\par In order to compile a sample of blazar candidates, we cross-matched the position of all the \textit{Swift} X-ray sources listed in the \textit{Swift} Serendipitous Survey in deep XRT GRB Fields catalog \citep{Swift} with a number of radio catalogs - the NRAO VLA Sky Survey  \citep[NVSS,][]{Condon} and the Faint Images of the Radio Sky at Twenty-cm \citep[FIRST,][]{becker_first}. After that, we restricted ourselves to the fields covered by the Sloan Digital Sky Survey (SDSS) DR14 \citep{sdss14}. This sample was then used to identify new blazars and build their X-ray LogN-LogS, down to an X-ray flux density of a few 10$^{-15}$ erg cm$^{-2}$ s$^{-1}$, and their radio LogN-logS, with fluxes down to approximately 10 mJy, well below the flux limit of previous complete blazar surveys. 

\par This paper is organized as follows: in Section \ref{finding} we describe the method used to obtain the possible sources and the cross matching with the SDSS DR14 database; in Section \ref{catalog} we present the catalog and its properties; in Section \ref{logns_radio} we present our radio LogN-LogS, whereas in Section \ref{logns_x} we build our X-ray LogN-LogS plot and compare it with other studies. In Section \ref{fin} we discuss our conclusions.

\section{Finding blazar candidates}\label{finding}

The \textit{Swift} satellite is a multi-frequency rapid response GRB space observatory. It carries three instruments on board: the  Burst Alert Telescope (BAT), sensible in the 15-150 keV band; the X-ray Telescope (XRT), sensible in the 0.2-10 keV band; and the UV and Optical Telescope (UVOT). 
\par As of today, \textit{Swift} discovered well over 1,000 GRBs, a good fraction of which were observed with XRT and UVOT to monitor the decay of GRB afterglows for several days. \citet{Swift} merged all the XRT images centered on GRBs observed from January 2005 to December 2008 to obtain long or very long exposures, from $\approx$ 10,000 to over one million seconds, with the sensitivity of the deepest images reaching $\approx 10^{-15}$ erg cm$^{-2}$ s$^{-1}$ in the soft X-ray band (0.5-2 keV). The catalog of point sources detected in these deep GRBs exposures includes more than 9,000 sources and can be accessed online at the SSDC website\footnote{http://www.ssdc.asi.it/xrtgrbdeep\_cat/}.
\par The earliest efforts to produce blazar samples involved searching large X-ray or radio surveys, following up with optical identification of the sources. However, with the increasing size of the catalogs, optical follow-up is demanding more and more telescope time, down to unmanageable levels.  Since one of the key features of blazars is that their emission covers the entire 
electromagnetic spectrum, to reduce the number of candidates, we first search for radio counterparts of X-ray sources, as in this way only objects that emit in a broad range of 
wavelengths are selected. 
\par We cross-matched the X-ray sources found in the \textit{Swift} Serendipitous Survey in deep XRT GRB Fields with radio catalogs such as the NVSS and the FIRST. The radius for this initial matching was 12 arcsec, somewhat larger than the typical XRT error circle of approximately five arcsec \citep{Swifterror} and the radio catalogs uncertainties to take into account that many of our X-ray sources are very faint and discovered in deep images and to avoid missing slightly radio extended objects and very faint radio sources. We obtained 125 X-ray/radio associations, which were then searched for optical counterparts in the SDSS DR14 within 12 arcsec from the X-ray position. Due to this relatively large area, we found that the 125 X-ray radio associations match 298 optical sources, some of them with several multiple optical counterparts (MOCs). We obtained positions, positional errors, magnitudes (\textit{ugriz}) and redshift (when available) for these sources.

\subsection{Source association}

In order to associate the best optical counterpart to all radio-X-ray candidates with MOCs, we implemented the likelihood ratio technique (LR) \citep{stat1,stat2} to estimate the probability 
that each optical object is the true counterpart to the X-ray source. Assuming that the XRT position errors are gaussian, the LR for each source is

\be
LR = \frac{Q(\leq m) e^{-\frac{r^2} {2}}} {2 \pi \sigma_{ox}^2 n(\leq  m)}
\ee  

where $Q(\leq m)$ is the a priori probability that a ``true'' optical counterpart brighter than the magnitude limit exists in the association, $n(\leq  m)$ is local surface density of  objects brighter than the candidate,  $\sigma_{ox} = \sqrt{ \sigma_{x}^2 + \sigma_{o}^2} $ is the total (X-ray+optical) positional uncertainty and $r$ is the ``normalized distance'', $r^{2} = 2 (\frac{\Delta_{ox}}{\sigma_{ox}})^2$, and $\Delta_{ox}$ is the actual distance between the X-ray and optical positions.

\par For simplicity, we set $Q(\leq m)=1$ in this work, assuming that the true optical counterpart always exist and it is above the magnitude limit. To properly calculate $\sigma_x$, we consider the results of \citet{Swifterror} and defined $r_{95}=2\sigma_x$ as 

\be
r_{95} = \sqrt{(5'')^2+{r_{stat}}^2} 
\ee

where, for each source,

\be
r_{stat} = 22.63'' {N_{ph}}^{-0.48}
\ee

with $N_{ph}$ is the number of ``effective counts'' in the full band,  i.e. the total number of counts between 0.2 and 10 keV after the subtraction of the average background value.
\par We can compute $n(\leq m)$ within a circle of radius $\sigma_{ox}$,

\be
n(\leq  m) =  \frac{N(\leq  m)}{4 \pi \sigma_{ox}^2}
\ee

with $N(\leq m)$ is the total number of sources with magnitude less or equal to that of the candidate. Then, we can calculate the LR as 

\be
 LR=\frac{2\exp(-(\frac{\Delta_{ox}}{\sigma_{ox}})^2)}{N(\leq m)}
\ee

 \par We computed the LR for each potential optical counterpart of each X-ray source in the sample and selected the ones with the highest values to build the best LR sample. We verified the reliability of the method by individually inspecting every source using the SSDC tools\footnote{tools.ssdc.asi.it}, and NED online services. We found that:

\begin{enumerate}
 \item two were spurious associations with SDSS;
 \item one was a wrong radio X-ray association;
 \item four don't have a clear optical counterpart (there are bright sources very near each other in the SDSS field);
 \item two were probably spurious detections on \textit{Swift} deep GRB fields (not detected in the full band);
 \item some sources were not blazars (radio extended, spiral galaxies or radio galaxies), so we removed them from the final sample.
\end{enumerate}

It is important to note that the source with a wrong radio X-ray association and the two objects with spurious SDSS associations have $LR\approx0$, which also confirms the reliability of the method used to select the candidates.

We underline that the likelihood method was used to assess the probability of each optical counterpart be the true counterpart of the X-ray source. Since there were no multiple radio X-ray matches, there was no need to use the likelihood method to classify them. Furthermore, each radio X-ray association has been visually inspected and verified by the authors.

\section{The Catalog}\label{catalog}

\begin{figure}
	\includegraphics[width=\columnwidth]{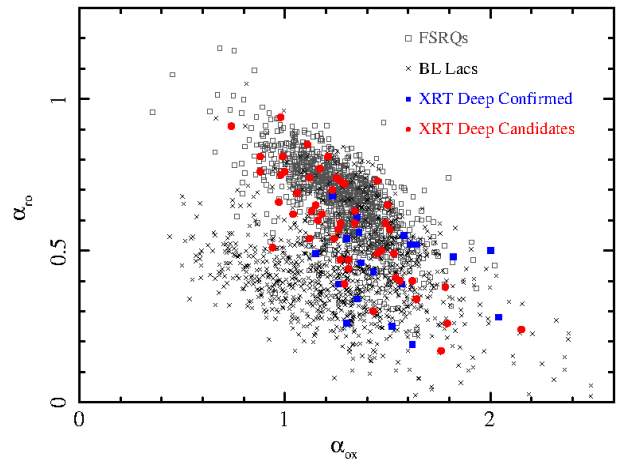}
    \caption{The \aro - \aox distribution for candidates blazar in our sample (large purple stars and blue crosses) superimposed to the one based on the blazars included in the BZCAT5 catalog (small circles).}
     \label{aroaox}
   \end{figure}

After removing the sources that either were clearly not blazars or had bad counterparts, the sample includes 62 good X-ray radio associations with an optical counterpart. We present these sources in Table \ref{total}, where
{\it column 1}: name from \textit{Swift} Serendipitous Survey in deep XRT GRB Fields catalog;
{\it column 2}: Right Ascension (J2000) from SDSS;
{\it column 3}: Declination (J2000) from SDSS;
{\it column 4}: Radio flux at 1.4GHz;
{\it column 5}: X-ray flux in the 0.5-2 KeV band;
{\it column 6}: SDSS redshift (when measured);
{\it column 7}: logarithm in base 10 of the synchrotron peak \nupeak;
{\it column 8}: logarithm in base 10 of $\nu F_\nu$;
{\it column 9}: the LR value;
{\it column 10}: classification.   

\par  We used a third order polynomial fit to calculate the synchrotron peak \nupeak and we found 1 HSP blazars, although it has a high uncertainty in the determination of \nupeak due to lack of data. We used the SSDC SED tool to fit the data, which, unfortunately, at present does not estimate the \nupeak error.

Of the 62 sources, 19 were found to be blazars based on their available optical spectra, and, among these, 8 are BL Lacs and 11 FSRQs. For the classification in Table \ref{total} we followed the one used in BZCAT5, that is BZB for BL Lac objects, BZQ for FSRQs in case of confirmed blazars, and BZG when the galaxy is clearly dominating the optical emission with respect to the nucleus. We classify the object as Candidate in the remaining cases. 

If we consider the confirmed blazars only, the fraction of HSPs is $\sim 5\%$ $(i.e. 1/19)$. The completion of the identification of the candidates is necessary to proper estimate the fraction of HPSs in the overall sample. However, we would like to note that the fraction of HSPs in an X-ray selected sample, which also has a radio cut such as ours, depends very much on the X-ray flux. In fact, most of HPSs are BL Lac objects; then, as shown in Fig. \ref{lognscompare}, simulations predict that at low X-ray fluxes the ratio between FSRQs and BL Lacs changes, with FSRQs becoming the main blazar subclass, therefore implying a possible change also in the percentage of HSPs in a given sample. Future studies will analyse this in more detail.

\par The distribution of the slopes between radio-optical and optical-X-ray for our sample superimposed to the one in the BZCAT5 catalog is shown in Figure \ref{aroaox}, where $\alpha_{\nu_1\nu_2}$ is defined as:
\begin{equation}
 \alpha_{\nu_1\nu_2}=-\frac{\log(f_1/f_2)}{\log(\nu_1/\nu_2)}.
\end{equation}
We converted our radio flux from $1.4$ GHz to 5GHz using a power law with index $-0.25$; the X-ray frequency is chosen at 1 KeV. Our sample falls well on the region populated by the BZCAT5 blazars.

We noticed that in Fig. \ref{aroaox} the area defined by \aox$>1.3$ and \aro$<0.6$ is more crowded with candidates. Two reasons can account for the presence of this tail in the distribution of our sources at the right-lower bottom on the plane \aox-\aro, namely: i) the area is defined by the intersection of the two branches of LPSs and HSPs, therefore it is populated mainly by sources having \nupeak at intermediate frequencies, therefore the density of sources in this region of the plot simply reflects the statistical occurrence of ISPs; ii) possible contamination of thermal radiation for weak X-ray sources that would affect their \aro and \aox indices. More specifically, in the case of BL Lacs, the contribution from host galaxy light in the optical will cause the \aro to decrease to lower values, and in the meantime \aox to move to greater values, whereas in the case of FSRQs, the contribution of the Big Blue Bump in the optical will also make \aro to decrease, while we infer that at the same time \aox will probably remain almost constant, as the observed relation between the X-ray thermal radiation in the corona of quasars, and the optical/UV emission from the disk can be described by the a well-established anti-correlation between the \aox and the UV luminosity \citep[see e.g.][]{vagn10, LR2016, CL2018}. However, we underline that these heuristic arguments are just a starting point for a proper discussion, and they could be investigated properly only by dedicated simulations of X-ray selected samples at low X-ray fluxes, and confirmed later with future studies.

We calculated the LogN-LogS in the radio band at 5GHz and in the X-ray band using our final sample, reaching very deep fluxes. We will present these new LogN-LogS in the following sections, moving from the lowest (radio) to the highest energy (X-ray).

\section{The radio LogN-LogS}\label{logns_radio}

\par The LogN-LogS of a population of sources in a given energy band can be used to estimate the emission in other parts of the electromagnetic spectrum, once the flux ratio in the two bands, or even better, the overall energy distribution, is known. 

We used our new blazar sample to estimate the radio LogN-LogS of blazars with fluxes down to 10 mJy. 

\begin{figure}
 \includegraphics[width=\columnwidth]{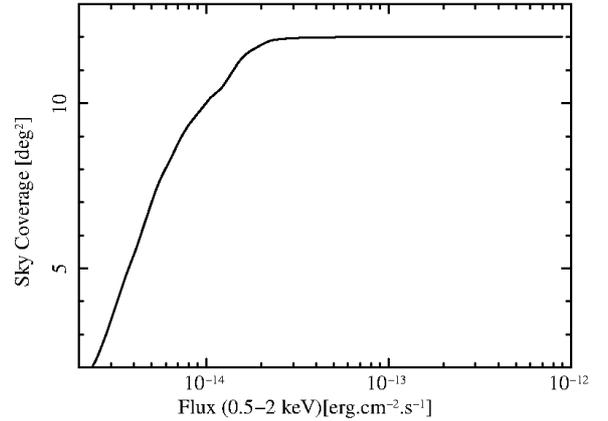}
 \caption{Sky coverage of the survey in terms of the X-ray flux for our sample of faint blazars.}
 \label{coverage}
   \end{figure}

\par We show in Fig. \ref{rlogns} our results together with calculations from previous surveys (see Appendix \ref{presur} for details). Since the radio surveys considered in this context have been carried out at different observing frequencies (1.4, 2.7 and 5 GHz), we converted all flux densities to a common band. We selected 5 GHz as the reference frequency and we apply flux conversions assuming as above a spectral slope $\alpha_r = 0.25$ ($f_\nu \sim \nu^{- \alpha_r}$) which is approximately equal to the average value in all the considered samples. 

The completeness limit of the NVSS is $\sim 2.5$mJy whereas the detection limit is $\sim 1$mJy over most of the FIRST survey. The sensitivity of SUMSS is similar to that of the NVSS. This implies a quite flat radio sky coverage over our GRB fields with a radio cut at  $\sim 2.5$mJy. As consequence, we take into account the X-ray sky coverage to calculate the radio logN-logS, as we are not sampling the X-ray sky homogeneously over the different fields.

 We used the \textit{Swift} sky coverage (see Figure \ref{coverage}) \footnote{The ``sky coverage'' defines the solid angle of the sky covered by a survey to a given flux limit, as a function of the flux.} for the GRB fields covered by SDSS photometry (S. Puccetti, private communication) in order to calculate the counts in the final sample, plotted in Fig \ref{rlogns} as red filled circles (color figure available online only). 
 We considered that the sky coverage in Fig. \ref{coverage} points out that faint objects with flux around 10$^{-14}$ erg cm$^{-2}$ s$^{-1}$ could be detected only in $\sim$ 8 square degrees, whereas objects one order of magnitude brighter (i.e. flux $\sim$  10$^{-13}$ erg cm$^{-2}$ s$^{-1}$) could be detected in $\sim$ 10 square degrees. 
 Taking this into account, we followed the method used for the One Jansky ASDC-RASS-NVSS blazar sample \citep{giommicmb} to calculate the radio counts for our final sample at 1.4 GHz. In particular, for each source we used the radio flux to count the object in the corresponding bin of radio flux density, and the X-ray flux to estimate the area covered by the survey from the X-ray sky coverage. Then, we converted the flux densities to 5 GHz to obtain the final logN-logS, assuming a spectral slope of 0.25 as we did for the other surveys. 
Given the small area of the \textit{Swift} survey ($\approx 12$ square degrees), and the low space density of blazars, our sample can probe only the faint tail of the radio LogN-LogS.
\par Considering the overall derived counts N($>$S) from the different surveys, they are consistent with a broken powerlaw described as $S^{-1.66}$ for fluxes $S > 10~mJy$, with a break at $S \approx 10~mJy$. We have to note that the slope below the break cannot be estimated accurately as the optical spectra available allows us to calculate only lower limits to the density of blazars: however, we underline that for fluxes $S <10~mJy$ a slope flattening is required also in order to avoid that the predicted blazar space density exceeds the observed total density of radio sources at a few mJy (e.g. NVSS, FIRST) which we plot in Fig. \ref{rlogns} as upper limits.  We chose to model the slope below the break as $S^{-0.9}$, since $-0.9$ is the average slope of the logN-logS of radio quiet AGN in the two flux decades below the break \citep{Rosati02,Moretti03} and we found that it is consistent with the available constraints. Our results are consistent with previous work on this topic \citep{giommicmb}, that reported a good fit to previously known data as $S^{-1.62}$ for fluxes $S > 15~mJy$ and pointed out for the presence of a break below 15 mJy.

\begin{figure}
	\includegraphics[width=\columnwidth]{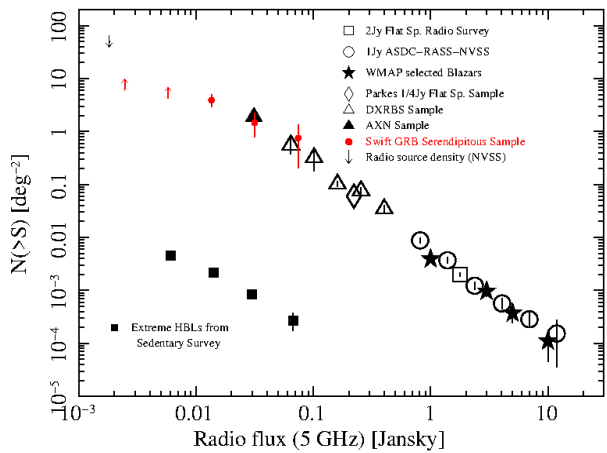}
    \caption{The radio (5 GHz) LogN-LogS of candidate blazars that extends to fluxes down to 10 mJy the blazar radio LogN-LogS built combining several radio and multi-frequency surveys. We show counts from each survey using different symbols (see text for details). The filled squares in the lower left part represent for comparison the radio LogN-LogS of extreme HBL BL Lacs from the Sedentary Survey.} 
     \label{rlogns}
   \end{figure}

However, we must consider the possibility that our sample is still contaminated within a certain amount by other non-thermal AGN characterized by steep radio spectrum, such as radio galaxies and steep spectrum quasars, with respect to blazars that show flat radio spectral slopes. Further multifrequency data, especially in the radio band and optical spectra, are needed to get conclusive classifications for our candidates.

Nonetheless, we would like to underline that at the lowest fluxes we could also start missing blazar identifications not only for the X-ray sky coverage but also given the radio cut and our request to have an optical counterpart in SDSS data. In fact, for example, it has been estimated that in case of SDSS-DR10, objects from the FIRST survey have $\sim 30\%$ of optical identifications at SDSS magnitude limit ($m_{V} \sim 23$).
Therefore, we are using lower limits on the density of blazars for the logN-logS points at the lowest fluxes in Fig. \ref{rlogns}, just in order to make more evident to the reader that the possibility of missing objects at these fluxes.

\section{The X-ray LogN-LogS}\label{logns_x}

\par
\begin{figure}
	\includegraphics[angle=-90, width=\columnwidth]{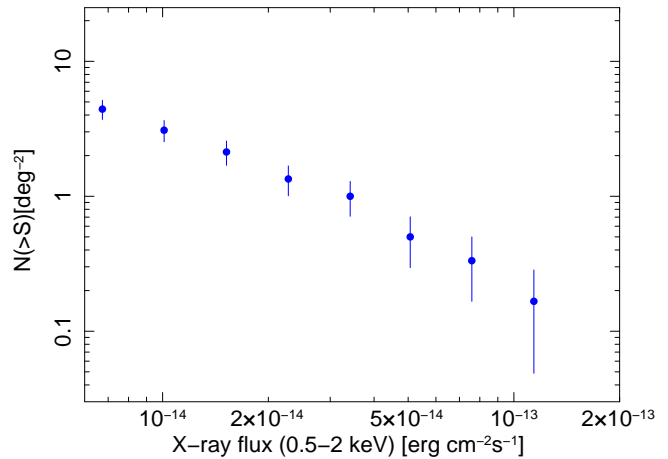}
    \caption{Total number density of the sources in our sample versus their X-ray fluxes.}
     \label{lognstot}
   \end{figure}

 The X-ray Log N-Log S for our sample of candidate blazars is shown in Figure \ref{lognstot}. It reaches very faint fluxes (below $10^{-14}$ erg cm$^{-2}$s$^{-1}$ in the 0.5-2.0 keV band) and it is therefore by far the deepest to date. We calculated the upper limits in Fig. \ref{lognstot} assuming that \textit{all} our candidates are blazars, whereas filled points represent actual number density for the confirmed blazars in the sample, i.e. confirmed on the basis of their optical spectra.

\par It is useful to compare the LogN-LogS of our sample with previously published logN-logS of blazars. One of the more recent estimations was done by \citet{giommixgamma}, who used a Monte Carlo simulated X-ray flux limited catalog to estimate the number counts of different types of faint blazars, showing that it agrees very well with previous estimates at bright fluxes. To compare our results with this one, it is necessary to rescale fluxes in our soft band (0.5-2 keV) to match the one used in the paper (0.3-3.5 keV) mentioned above.  Assuming an energy distribution with a  power law index of 0.9, the correction factor is 1.87.

\begin{figure}
	\includegraphics[width=\columnwidth]{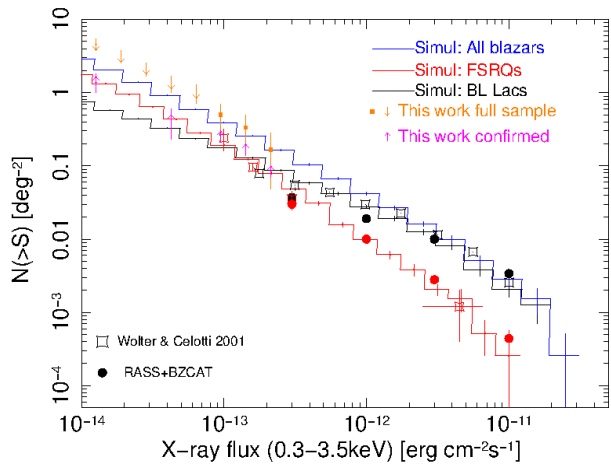}
    \caption{X-ray LogN-LogS of our final \textit{Swift} sample compared to the one of \citet{giommixgamma}. Orange arrows and filled squares correspond to the full sample (candidates and confirmed blazars), purple arrows are the spectroscopically confirmed blazars only. While our full sample overestimates the number of sources for faint fluxes ($\lesssim 10^{-13}$ erg cm$^{-2}$s$^{-1}$), the number of confirmed blazars is still below the simulated counts. See text for details. }
     \label{lognscompare}
   \end{figure}

\par As we can see from Figure \ref{lognscompare}, we are in good agreement at high fluxes, whereas our upper limits (orange downward arrows in the color figure available only online) estimated assuming that {\it all} our candidates are blazars overestimate the counts of \citet{giommixgamma} at the faint end. This is very likely due to the fact that not all of our sources will turn out to be confirmed as blazars.
On the other hand, if we build the LogN-LogS using only for the confirmed blazars (19 out of the 62, shown as purple arrows in the color figure available only online), we see that we are below the number predicted by the simulations. Therefore, both our number counts and the simulations of \citet{giommixgamma} could be in agreement after the removal of the non-blazar sources of our sample.

\par With respect to \citet{Wol01b}, our sample can probe populations of sources 10 times weaker in the X-ray flux, thus allowing for a more detailed comparison between data and simulated counts. In particular, as shown in Figure \ref{lognscompare}, simulations of  \citet{giommixgamma} predict the inversion of the population density of BL Lacs and FSRQs at lower fluxes. So far it was not possible to investigate this outcome from the simulations as the inversion was expected to happen at fluxes not reached by previous available samples. Therefore, we highlight that the complete classification of our candidates by means of optical spectra will be crucial to explore the properties of blazars at the really faint end of their Log N-Log S.  

\par Given the small area covered by \textit{Swift}, however, our sample can be used to study only populations below $10^{-13}$ erg cm$^{-2}$s$^{-1}$. 
On the contrary, data from the RASS-BZCAT sample \citep[see for details][]{giommixgamma} cover a sufficiently large area of the sky to investigate the numbers of FSRQs at high fluxes ($> 10^{-12}$ erg cm$^{-2}$s$^{-1}$), and it can be used to complement information from our sample.

 Furthermore, future studies of X-ray selected samples of faint blazars with eROSITA \citep{erosita} will shed more light on the blazar population at low X-ray fluxes.

\begin{table*}
\label{total}
\centering
\scriptsize
\caption{   {Catalog of faint blazars and blazar candidates.}}
\begin{tabular}{lcccccccccc}
\hline
NAME         &    {SDSS RA}         &    {SDSS DEC}      & Flux$_{1.4GHz}$   & Flux$_{0.5-2keV}$                 & z  & Log($\nu_p)$ & Log($\nu F_{\nu}$) &    {LR} & Class \\
             &      {(J2000, deg)}        &       {(J2000, deg)}      & (mJy)  &    {(units of $10^{-14}$)} &    &              &                    &     &       \\
    &        &       &          &   {(erg cm$^{-2}$s$^{-1}$)} &    &              &                    &     &       \\       
\hline 

   SWIFTFTJ001252.7$+$3241.6 & 3.22250 & 32.69389 & 8.20 & 3.44 & - &   12.5$^*$ &    -12.5$^*$ & 0.012 & Candidate \\  
   SWIFTFTJ005503.5$+$1408.0 & 13.76417 & 14.13500 & 99.9 & 4.10 & 1.666 &    13.1 &    -12.5 & 1.852 & BZQ \\
   SWIFTFTJ005514.7$+$1407.4 & 13.81083 & 14.12417 & 19.9 & 2.13 & - &    13.5 &    -12.7 & 1.872 & Candidate \\
   SWIFTFTJ012320.3$+$3812.9 & 20.83458 & 38.21583 & 6.70 & 0.345 & - &    12.1$^*$ &    -12.3$^*$ & 0.980 & Candidate \\
   SWIFTFTJ020726.5$+$0022.5 & 31.86042 & 0.37611 & 61.8 & 0.671 & - &    $^*$ &    $^*$ & 0.594 & Candidate \\
   SWIFTFTJ021305.6$-$0219.4 & 33.27333 & -2.32333 & 32.2 & 1.94 & 1.670 &    12.4$^*$ &    -12.6$^*$ & 1.723 & BZQ \\
   SWIFTFTJ021307.9$-$0212.3 & 33.28416 & -2.20528 & 11.4 & 0.935 & - &    $^*$ &    $^*$ & 0.198 & Candidate \\
   SWIFTFTJ024446.7$-$0210.0 & 41.19417 & -2.16806 & 1.60 & 2.15 & 2.012 &    12.5$^*$ &    -12.6$^*$ & 1.373 & BZQ \\
   SWIFTFTJ033630.2$+$1723.2 & 54.12542 & 17.38778 & 19.9 & 2.66 & - &    13.3$^*$ &    -13.1$^*$ & 0.546 & Candidate \\
   SWIFTFTJ054613.3$+$6410.5 & 86.55625 & 64.17611 & 12.6 & 0.490 & - &    $^*$ &    $^*$ & 1.715 & Candidate \\
   SWIFTFTJ062257.9$-$0109.4 & 95.74084 & -1.15611 & 241.0 & 3.45 & - &    12.9 & -12.7 & 0.074 & Candidate \\
   SWIFTFTJ075144.8$+$3107.9 & 117.93667 & 31.13222 & 24.2 & 0.362 & - &    13.0$^*$ &    -12.8$^*$ & 1.768 & Candidate \\
   SWIFTFTJ084803.7$+$1338.8 & 132.01543 & 13.64750 & 1.90 & 0.367 & - &    12.8$^*$ &    -12.7$^*$ & 0.911 & Candidate \\
   SWIFTFTJ084842.8$+$1336.3 & 132.17917 & 13.60639 & 7.60 & 1.43 & - &    12.8$^*$ &    -12.5$^*$ & 0.355 & Candidate \\
   SWIFTFTJ085542.7$+$1103.2 & 133.92792 & 11.05389 & 14.8 & 9.19 & 0.300 &    15.1$^*$$^1$ &    -12.5$^*$ & 1.727 & BZG \\
   SWIFTFTJ090602.1$+$3512.1 & 136.50792 & 35.20417 & 3.20 & 0.423 & - &    12.8$^*$ &    -12.3$^*$ & 0.000 & Candidate \\
   SWIFTFTJ090936.0$+$4547.5 & 137.39917 & 45.79306 & 24.7 & 0.547 & 0.321 &    13.8$^*$ &    -12.8 & 1.126 & BZG \\
   SWIFTFTJ090954.4$+$4544.3 & 137.47708 & 45.73667 & 57.8 & 4.12 & - & 12.9 & -13.0 & 0.003 & Candidate \\
   SWIFTFTJ091036.3$+$4537.2 & 137.65042 & 45.62083 & 2.70 & 0.716 & - &    13.6$^*$ &    -13.0$^*$ & 0.442 & Candidate \\
   SWIFTFTJ093045.1$+$1659.4 & 142.68918 & 16.99139 & 72.5 & 0.280 & 0.177 &    13.5$^*$ &    -12.5$^*$ & 1.062 & BZG \\
   SWIFTFTJ093750.9$+$1536.5 & 144.46251 & 15.60972 & 4.60 & 0.649 & - &    $^*$ &    $^*$ & 1.846 & Candidate \\
   SWIFTFTJ101433.3$+$4306.0 & 153.63916 & 43.10194 & 4.90 & 1.52 & 1.684 &    $^*$ &    $^*$ & 0.592 & BZQ \\
   SWIFTFTJ101609.4$+$4336.2 & 154.03917 & 43.60278 & 3.70 & 11.7 & 0.587 &    13.4$^*$ &    -12.6$^*$ & 0.762 & BZQ \\
   SWIFTFTJ101700.5$+$4328.3 & 154.25249 & 43.47195 & 5.90 & 0.757 & - &    12.5$^*$ &    -12.7$^*$ & 1.458 & Candidate \\
   SWIFTFTJ101727.4$+$4329.0 & 154.36417 & 43.48444 & 197.1 & 5.12 & 1.175 & 12.7 &    -12.5 & 1.745 & BZQ \\
   SWIFTFTJ110035.7$+$5148.2 & 165.14958 & 51.80333 & 4.20 & 1.47 & - &    14.0$^*$ &    -13.0$^*$ & 0.483 & Candidate \\
   SWIFTFTJ114449.6$+$5953.3 & 176.20667 & 59.88861 & 6.00 & 0.913 & - &    $^*$ &    $^*$ & 1.660 & Candidate \\
   SWIFTFTJ115036.9$+$5707.6 & 177.65500 & 57.12806 & 17.2 & 0.733 & 0.117 &    14.3$^*$ &    -13.6$^*$ & 0.150 & BZG \\
   SWIFTFTJ120512.5$+$4007.0 & 181.30125 & 40.11583 & 6.00 & 0.431 & 2.434 &    13.6$^*$ &    -13.1$^*$ & 0.455 & BZQ \\
   SWIFTFTJ121012.4$+$3959.0 & 182.55292 & 39.98417 & 4.50 & 0.512 & 0.562 &    $^*$ &    $^*$ & 1.297 & BZG \\
   SWIFTFTJ123405.8$+$2102.4 & 188.52415 & 21.04028 & 1.20 & 1.29 & - &    13.1$^*$ &    -13.0$^*$ & 1.860 & Candidate \\
   SWIFTFTJ130358.6$+$4110.1 & 195.99542 & 41.16972 & 1.90 & 0.604 & 1.212 &    $^*$ &    $^*$ & 0.081 & BZQ \\
   SWIFTFTJ131215.9$+$6200.9 & 198.06792 & 62.01694 & 16.3 & 2.82 & - &    13.6$^*$ &    -13.0$^*$ & 0.137 & Candidate \\
   SWIFTFTJ132332.3$+$4043.5 & 200.88834 & 40.72722 & 28.6 & 3.00 & - &    13.3$^*$ &    -13.2$^*$ & 0.008 & Candidate \\
   SWIFTFTJ132928.6$+$4230.7 & 202.37000 & 42.51361 & 30.5 & 0.942 & 1.597 &    13.0$^*$ &    -12.9$^*$ & 0.940 & BZQ \\
   SWIFTFTJ133128.5$+$4209.7 & 202.86958 & 42.16195 & 1.30 & 1.17 & 0.939 & 12.8 & -12.4 & 0.710 & BZQ \\
   SWIFTFTJ133201.5$+$3458.9 & 203.00667 & 34.98361 & 0.80 & 0.0997 & - &    13.6 &    -13.2 & 1.000 & Candidate \\
   SWIFTFTJ134931.4$+$0732.7 & 207.38167 & 7.54528 & 5.00 & 1.18 & - &    13.9$^*$ &    -12.6$^*$ & 0.871 & Candidate \\
   SWIFTFTJ141144.6$+$1655.2 & 212.93875 & 16.92194 & 1.60 & 2.07 & 0.615 &    $^*$ &    $^*$ & 0.00 & BZB \\
   SWIFTFTJ143133.6$+$3628.0 & 217.88792 & 36.46611 & 4.70 & 0.183 & - &    12.7$^*$ &    -13.0$^*$ & 0.035 & Candidate \\
   SWIFTFTJ143733.7$+$2743.3 & 219.39041 & 27.72278 & 15.7 & 0.208 & 0.310 &    14.0$^*$ &    -12.8$^*$ & 1.709 & BZG \\
   SWIFTFTJ144021.7$-$0004.6 & 220.09041 & -0.07722 & 16.7 & 0.440 & - &    13.5$^*$ &    -13.4$^*$ & 0.999 & Candidate \\
   SWIFTFTJ144050.1$+$3333.8 & 220.20876 & 33.56389 & 7.00 & 4.20 & 1.777 &    12.6$^*$ &    -12.2$^*$ & 1.729 & BZQ \\
   SWIFTFTJ144615.4$+$5437.0 & 221.56459 & 54.61861 & 1.00 & 2.37 & - &    14.2$^*$ &    -13.4$^*$ & 1.055 & Candidate \\
   SWIFTFTJ151338.1$+$3048.2 & 228.40958 & 30.80333 & 3.40 & 4.61 & - &    14.0$^*$ &    -12.6$^*$ & 0.279 & Candidate \\
   SWIFTFTJ151526.3$+$4424.0 & 228.85918 & 44.40250 & 15.8 & 0.647 & - &    13.2$^*$ &    -12.9$^*$ & 0.109 & Candidate \\
   SWIFTFTJ153133.2$+$6327.8 & 232.88792 & 63.46305 & 64.0 & 0.515 & - &    $^*$ &    $^*$ & 1.453 & Candidate \\
   SWIFTFTJ153143.3$+$0020.3 & 232.93126 & 0.33639 & 37.3 & 0.911 & - &    13.1$^*$ &    -12.9$^*$ & 0.003 & Candidate \\
   SWIFTFTJ154059.3$+$6205.0 & 235.24791 & 62.08278 & 1.40 & 1.20 & - &    13.3 &    -12.1 & 1.139 & Candidate \\
   SWIFTFTJ155127.7$+$4447.4 & 237.86667 & 44.78944 & 6.30 & 0.663 & - &    12.0$^*$ &    -12.9$^*$ & 0.527 & Candidate \\
   SWIFTFTJ165845.5$+$1220.4 & 254.68916 & 12.34167 & 77.5 & 1.81 & - &    13.2 &    -12.9 & 1.765 & Candidate \\
   SWIFTFTJ183230.6$+$4230.2 & 278.12668 & 42.50389 & 25.0 & 14.0 & - &    13.5 & -12.4 & 0.377 & Candidate \\
   SWIFTFTJ213156.7$+$0246.0 & 322.98584 & 2.76861 & 15.4 & 0.415 & 0.387 &    $^*$ &    $^*$ & 1.850 & BZG \\
   SWIFTFTJ215415.6$+$1652.6 & 328.56543 & 16.87611 & 17.1 & 5.57 & - &    12.8$^*$ &    -12.8$^*$ & 0.987 & Candidate \\
   SWIFTFTJ215436.0$+$1653.2 & 328.65082 & 16.88667 & 5.20 & 0.448 & - &    13.8$^*$ &    -13.1$^*$ & 0.230 & Candidate \\
   SWIFTFTJ220849.1$+$0652.3 & 332.20416 & 6.87278 & 229.4 & 7.73 & - &    14.5 &    -12.3 & 1.048 & Candidate \\
   SWIFTFTJ222507.4$-$0223.4 & 336.28168 & -2.39111 & 39.1 & 2.00 & - &    $^*$ &    $^*$ & 1.079 & Candidate \\
   SWIFTFTJ224220.2$+$2346.8 & 340.58417 & 23.78167 & 25.2 & 0.356 & - &    12.7$^*$ &    -12.9$^*$ & 1.761 & Candidate \\
   SWIFTFTJ230410.9$+$0357.4 & 346.04459 & 3.95778 & 9.40 & 1.69 & - &    13.6$^*$ &    -12.$^*$ & 1.625 & Candidate \\
   SWIFTFTJ232236.8$+$0538.9 & 350.65335 & 5.64889 & 4.70 & 1.03 & - &    13.8$^*$ &    -12.9$^*$ & 1.997 & Candidate \\
   SWIFTFTJ232311.1$+$0543.1 & 350.79541 & 5.71917 & 9.10 & 0.471 & - &    12.4$^*$ &    -12.6$^*$ & 0.323 & Candidate \\
   SWIFTFTJ234732.3$+$0016.9 & 356.88541 & 0.28361 & 1.20 & 0.666 & - &    13.0 &    -12.9 & 0.057 & Candidate \\

\hline 
Notes: $^1$ HSP \\
~~~~~~~~~   {$^*$ Uncertain value}
\end{tabular}
\end{table*}

\section{Conclusions}\label{fin}

\par By using faint sources serendipitously detected in \textit{Swift} GRB fields and cross-matching them with radio surveys, we built a flux-limited catalog of blazar candidates down to very faint X-ray fluxes, a region that was lacking coverage in previous blazar research. Restricting ourselves to the area covered by the SDSS, we managed to obtain magnitudes and spectra (when available) for several of our sources. We used this sample of blazar candidates to calculate the radio LogN-LogS of Blazars with fluxes approximately down to 10 mJy. 
We were also able to estimate the soft X-ray LogN-LogS of blazars down to fluxes at least one order of magnitude fainter than previous works. A comparison with the expectations from Monte Carlo simulations of X-ray surveys indicates that our catalog is somewhat contaminated by sources other than blazars. However, the number counts of the spectroscopically confirmed sources still falls below the expected densities. Therefore, a complete optical follow-up is necessary to refine the sample. Moreover, high frequency radio observations are needed to measure the spectral slope for the candidates which have radio measurements at a single frequency. In fact, high frequency data can better evaluate the nuclear spectra, as the slopes at lower frequencies could be affected by the contribution from radio extended components in the jet. Useful data for this purpose could in principle be already present in the VLA archive as many radio observations are carried out at the VLA once a GRB explodes in order to catch its radio afterglow. Otherwise, the new JVLA \citep{jvla,Perley2009} and the upgraded ATCA \citep{mmATCA} are both promising to explore the spectral slope of the candidates as both these arrays have wider spectral radio bands. In particular, useful data will probably come from the on-going VLA Sky Survey (VLASS; Lacy et al. in preparation)\footnote{https://science.nrao.edu/science/surveys/vlass }. 

Completing the spectroscopical identification is crucial also because it will allow us to study the two main sub-populations of blazars, in particular regarding the inversion of their relative number density at low fluxes.
We stress that to date the cosmological evolution properties of blazars have been studied on a few samples with a sufficiently large size. So far, it has been established that FSRQ evolve positively, both in radio and X-ray selected samples, as showed by \citet{Wol01b}, with the first X-ray selected sample for FSRQs.
Results from literature indicate a higher evolution for X-ray selected quasars, although consistent at the $2 \sigma$ level \citep{Wol01b}.
Less clear is the trend among the BL Lac objects; however there seems to be a difference between the two classes: LSP BL Lacs show a slight positive evolution, consistent with no evolution at the $2\sigma$ level \citep{Sti91}, quite similar to radio selected FSRQs. HPS BL Lacs instead, show a negative evolution, more or less at the same $\sigma$ level, \citep{Rec00}, indicating that the X-ray bright objects are less luminous or less numerous at high redshifts. 

It has been shown that FSRQs evolve positively, whereas BL Lacs show no strong evolution also in the 15-55 KeV band, using an X-ray selected sample with data from the BAT instrument onboard \textit{Swift} \citep{addconcl2}. Instead, BL Lacs show positive evolution in case of $\gamma$-ray selected sample, with the relevant exception of low-luminosity HSP BL Lacs, which exhibit strong negative evolution \citep{addconcl1}. 
Recently, \citet{addconcl3} studied a sample of 26 high-redshift ($z>4$) radio-selected FSRQs, and found results in agreement with the predictions from the luminosity function derived on a radio-selected sample of FSRQs at lower redshifts \citep{addconcl4}, i.e. consistent with a peak in the space densities of FSRQs at $z \sim 2$, similar to what is found for radio-quiet QSOs. However, these findings on FSRQs are in tension with the results from the BAT X-ray selected sample, which on the contrary found a peak at $z \sim 4$ \citep{addconcl2}. To date, there is no clear explanation of this difference, and further investigation is needed, especially to constrain the evolution of sources at lower luminosity.

Therefore, completing the optical identifications for our candidates will be important in this context as we will be able to investigate blazar cosmological properties down to really faint fluxes, 10 times weaker than probed by previous works.

\par We have to note however - as also stated in \citet{Mignani2009} - that the currently available public optical surveys do not provide sufficient data and spectra to support a systematic X-ray source identification work. Our work was carried out using object lists matching SDSS optical sources. Unfortunately there is not an analogous of the SDSS photometric survey in the southern sky: in fact, the USNO-B1 \citep{USNOB} catalog is too shallow to search for optical counterparts for the \textit{Swift} radio-X associations lying in the southern sky. USNO-B1 limiting magnitude is $R \approx 20$, whereas SDSS limiting magnitude is deeper, $r' \sim 22$. This represents a severe limitation since the deepest flux limits reached by the \textit{Swift} survey requires of course similarly deeper optical catalogs to homogeneously sample the \aox-\aro parameter space. Surely, on-going and next generation surveys in the southern emisphere, such as VISTA  \citep{VISTA}, Pan-STARSS  \citep{Kaiser2010} and LSST \citep{Ivezic2012}, will provide useful data for optical identifications of faint high energy sources. 

\section*{Acknowledgements}
Part of this work is based on archival data, software or online services provided by the Space Science Data Center - ASI.
Additional data have been obtained from the NASA/IPAC Extragalactic Database (NED), and from the Sloan Digital Sky Survey SDSS-DR14 SkyServer.

 Funding for the Sloan Digital Sky Survey IV has been provided by the Alfred P. Sloan Foundation, the U.S. Department of Energy Office of Science, and the Participating Institutions. SDSS-IV acknowledges support and resources from the Center for High-Performance Computing at
the University of Utah. The SDSS web site is www.sdss.org.

SDSS-IV is managed by the Astrophysical Research Consortium for the 
Participating Institutions of the SDSS Collaboration including the 
Brazilian Participation Group, the Carnegie Institution for Science, 
Carnegie Mellon University, the Chilean Participation Group, the French Participation Group, Harvard-Smithsonian Center for Astrophysics, 
Instituto de Astrof\'isica de Canarias, The Johns Hopkins University, Kavli Institute for the Physics and Mathematics of the Universe (IPMU) / 
University of Tokyo, the Korean Participation Group, Lawrence Berkeley National Laboratory, Leibniz Institut f\"ur Astrophysik Potsdam (AIP),  Max-Planck-Institut f\"ur Astronomie (MPIA Heidelberg), Max-Planck-Institut f\"ur Astrophysik (MPA Garching), Max-Planck-Institut f\"ur Extraterrestrische Physik (MPE), National Astronomical Observatories of China, New Mexico State University, 
New York University, University of Notre Dame, Observat\'ario Nacional / MCTI, The Ohio State University, Pennsylvania State University, Shanghai Astronomical Observatory, United Kingdom Participation Group, Universidad Nacional Aut\'onoma de M\'exico, University of Arizona, University of Colorado Boulder, University of Oxford, University of Portsmouth, University of Utah, University of Virginia, University of Washington, University of Wisconsin, Vanderbilt University, and Yale University.

ST acknowledges financial support by Regione Lazio during the initial phase of this project. 
BF was supported by the CAPES-ICRANet program (BEX 14205-13-0) and by FAPERJ grants n. 202.687/2016 and 202.688/2016. 
PG acknowledges the support of the Technische Universit\"at M\"unchen - Institute for Advanced Studies, funded by the German Excellence Initiative (and the European Union Seventh Framework Programme under grant agreement no. 291763).

The authors would like to thank Bruno Sversut Arsioli for his help during an intermediate phase of the project,and Simonetta Puccetti for providing us the \textit{Swift} X-ray sky coverage.




\bibliographystyle{mnras} 

\begin{thebibliography}{}
\makeatletter
\relax
\def\mn@urlcharsother{\let\do\@makeother \do\$\do\&\do\#\do\^\do\_\do\%\do\~}
\def\mn@doi{\begingroup\mn@urlcharsother \@ifnextchar [ {\mn@doi@}
  {\mn@doi@[]}}
\def\mn@doi@[#1]#2{\def\@tempa{#1}\ifx\@tempa\@empty \href
  {http://dx.doi.org/#2} {doi:#2}\else \href {http://dx.doi.org/#2} {#1}\fi
  \endgroup}
\def\mn@eprint#1#2{\mn@eprint@#1:#2::\@nil}
\def\mn@eprint@arXiv#1{\href {http://arxiv.org/abs/#1} {{\tt arXiv:#1}}}
\def\mn@eprint@dblp#1{\href {http://dblp.uni-trier.de/rec/bibtex/#1.xml}
  {dblp:#1}}
\def\mn@eprint@#1:#2:#3:#4\@nil{\def\@tempa {#1}\def\@tempb {#2}\def\@tempc
  {#3}\ifx \@tempc \@empty \let \@tempc \@tempb \let \@tempb \@tempa \fi \ifx
  \@tempb \@empty \def\@tempb {arXiv}\fi \@ifundefined
  {mn@eprint@\@tempb}{\@tempb:\@tempc}{\expandafter \expandafter \csname
  mn@eprint@\@tempb\endcsname \expandafter{\@tempc}}}

\bibitem[\protect\citeauthoryear{{Abdo} et~al.,}{{Abdo}
  et~al.}{2010}]{abdosed10}
{Abdo} A.~A.,  et~al., 2010, \mn@doi [\apj] {10.1088/0004-637X/716/1/30}, \href
  {http://adsabs.harvard.edu/abs/2010ApJ...716...30A} {716, 30}

\bibitem[\protect\citeauthoryear{{Abolfathi} et~al.,}{{Abolfathi}
  et~al.}{2018}]{sdss14}
{Abolfathi} B.,  et~al., 2018, \mn@doi [\apjs] {10.3847/1538-4365/aa9e8a},
  \href {http://adsabs.harvard.edu/abs/2018ApJS..235...42A} {235, 42}

\bibitem[\protect\citeauthoryear{{Ackermann} et~al.,}{{Ackermann}
  et~al.}{2017}]{addrecent1}
{Ackermann} M.,  et~al., 2017, \mn@doi [\apjl] {10.3847/2041-8213/aa5fff},
  \href {http://adsabs.harvard.edu/abs/2017ApJ...837L...5A} {837, L5}

\bibitem[\protect\citeauthoryear{{Ajello} et~al.,}{{Ajello}
  et~al.}{2009}]{addconcl2}
{Ajello} M.,  et~al., 2009, \mn@doi [\apj] {10.1088/0004-637X/699/1/603}, \href
  {http://adsabs.harvard.edu/abs/2009ApJ...699..603A} {699, 603}

\bibitem[\protect\citeauthoryear{{Ajello} et~al.,}{{Ajello}
  et~al.}{2014}]{addconcl1}
{Ajello} M.,  et~al., 2014, \mn@doi [\apj] {10.1088/0004-637X/780/1/73}, \href
  {http://adsabs.harvard.edu/abs/2014ApJ...780...73A} {780, 73}

\bibitem[\protect\citeauthoryear{{Arsioli}, {Fraga}, {Giommi}, {Padovani}  \&
  {Marrese}}{{Arsioli} et~al.}{2015}]{1whsp}
{Arsioli} B.,  {Fraga} B.,  {Giommi} P.,  {Padovani} P.,   {Marrese} P.~M.,
  2015, \mn@doi [\aap] {10.1051/0004-6361/201424148}, \href
  {http://adsabs.harvard.edu/abs/2015A%26A...579A..34A} {579, A34}

\bibitem[\protect\citeauthoryear{{Becker}, {White}  \& {Helfand}}{{Becker}
  et~al.}{1995}]{becker_first}
{Becker} R.~H.,  {White} R.~L.,   {Helfand} D.~J.,  1995, \mn@doi [\apj]
  {10.1086/176166}, \href {http://adsabs.harvard.edu/abs/1995ApJ...450..559B}
  {450, 559}

\bibitem[\protect\citeauthoryear{Bennett et~al.,}{Bennett
  et~al.}{2003}]{bennett03}
Bennett C.~L.,  et~al., 2003, ApJS, 148, 97

\bibitem[\protect\citeauthoryear{{Blandford} \& {Rees}}{{Blandford} \&
  {Rees}}{1978}]{blandford}
{Blandford} R.~D.,  {Rees} M.~J.,  1978, \mn@doi [\physscr]
  {10.1088/0031-8949/17/3/020}, \href
  {http://adsabs.harvard.edu/abs/1978PhyS...17..265B} {17, 265}

\bibitem[\protect\citeauthoryear{{Boller}, {Freyberg}, {Tr{\"u}mper}, {Haberl},
  {Voges}  \& {Nandra}}{{Boller} et~al.}{2016}]{2rxs}
{Boller} T.,  {Freyberg} M.~J.,  {Tr{\"u}mper} J.,  {Haberl} F.,  {Voges} W.,
  {Nandra} K.,  2016, \mn@doi [\aap] {10.1051/0004-6361/201525648}, \href
  {http://adsabs.harvard.edu/abs/2016A\%26A...588A.103B} {588, A103}

\bibitem[\protect\citeauthoryear{{Caccianiga} et~al.,}{{Caccianiga}
  et~al.}{2019}]{addconcl3}
{Caccianiga} A.,  et~al., 2019, \mn@doi [\mnras] {10.1093/mnras/sty3526}, \href
  {http://adsabs.harvard.edu/abs/2019MNRAS.484..204C} {484, 204}

\bibitem[\protect\citeauthoryear{{Chandler} \& {Butler}}{{Chandler} \&
  {Butler}}{2014}]{jvla}
{Chandler} C.~J.,  {Butler} B.~J.,  2014, in Observatory Operations:
  Strategies, Processes, and Systems V. p. 914917, \mn@doi{10.1117/12.2057106}

\bibitem[\protect\citeauthoryear{{Chiaraluce}, {Vagnetti}, {Tombesi}  \&
  {Paolillo}}{{Chiaraluce} et~al.}{2018}]{CL2018}
{Chiaraluce} E.,  {Vagnetti} F.,  {Tombesi} F.,   {Paolillo} M.,  2018, \mn@doi
  [\aap] {10.1051/0004-6361/201833631}, \href
  {http://adsabs.harvard.edu/abs/2018A%26A...619A..95C} {619, A95}

\bibitem[\protect\citeauthoryear{{Condon}, {Cotton}, {Greisen}, {Yin},
  {Perley}, {Taylor}  \& {Broderick}}{{Condon} et~al.}{1998}]{Condon}
{Condon} J.~J.,  {Cotton} W.~D.,  {Greisen} E.~W.,  {Yin} Q.~F.,  {Perley}
  R.~A.,  {Taylor} G.~B.,   {Broderick} J.~J.,  1998, \mn@doi [\aj]
  {10.1086/300337}, \href {http://esoads.eso.org/abs/1998AJ....115.1693C} {115,
  1693}

\bibitem[\protect\citeauthoryear{{D'Abrusco}, {Massaro}, {Paggi}, {Smith},
  {Masetti}, {Landoni}  \& {Tosti}}{{D'Abrusco} et~al.}{2014}]{WGS2014}
{D'Abrusco} R.,  {Massaro} F.,  {Paggi} A.,  {Smith} H.~A.,  {Masetti} N.,
  {Landoni} M.,   {Tosti} G.,  2014, \mn@doi [\apjs]
  {10.1088/0067-0049/215/1/14}, \href
  {http://adsabs.harvard.edu/abs/2014ApJS..215...14D} {215, 14}

\bibitem[\protect\citeauthoryear{{Gehrels} et~al.,}{{Gehrels}
  et~al.}{2004}]{Swift1}
{Gehrels} N.,  et~al., 2004, \mn@doi [\apj] {10.1086/422091}, \href
  {http://adsabs.harvard.edu/abs/2004ApJ...611.1005G} {611, 1005}

\bibitem[\protect\citeauthoryear{{Gioia}, {Maccacaro}, {Schild}, {Wolter},
  {Stocke}, {Morris}  \& {Henry}}{{Gioia} et~al.}{1990}]{emss}
{Gioia} I.~M.,  {Maccacaro} T.,  {Schild} R.~E.,  {Wolter} A.,  {Stocke} J.~T.,
   {Morris} S.~L.,   {Henry} J.~P.,  1990, \mn@doi [\apjs] {10.1086/191426},
  \href {http://adsabs.harvard.edu/abs/1990ApJS...72..567G} {72, 567}

\bibitem[\protect\citeauthoryear{{Giommi} \& {Padovani}}{{Giommi} \&
  {Padovani}}{2015}]{giommixgamma}
{Giommi} P.,  {Padovani} P.,  2015, \mn@doi [\mnras] {10.1093/mnras/stv793},
  \href {http://adsabs.harvard.edu/abs/2015MNRAS.450.2404G} {450, 2404}

\bibitem[\protect\citeauthoryear{{Giommi}, {Menna}  \& {Padovani}}{{Giommi}
  et~al.}{1999}]{sedentary}
{Giommi} P.,  {Menna} M.~T.,   {Padovani} P.,  1999, \mn@doi [\mnras]
  {10.1046/j.1365-8711.1999.02942.x}, \href
  {http://adsabs.harvard.edu/abs/1999MNRAS.310..465G} {310, 465}

\bibitem[\protect\citeauthoryear{Giommi, Perri, Piranomonte  \&
  Padovani}{Giommi et~al.}{2002}]{Giommi02}
Giommi P.,  Perri M.,  Piranomonte S.,   Padovani P.,  2002, in Giommi P.,
  Massaro E.,   Palumbo G.,  eds, Blazar Astrophysics with {\it BeppoSAX} and
  Other Observatories. p.~123

\bibitem[\protect\citeauthoryear{Giommi, Piranomonte, Perri  \&
  Padovani}{Giommi et~al.}{2005}]{Gio05}
Giommi P.,  Piranomonte S.,  Perri M.,   Padovani P.,  2005, A\&A, 434, 385

\bibitem[\protect\citeauthoryear{{Giommi}, {Colafrancesco}, {Cavazzuti},
  {Perri}  \& {Pittori}}{{Giommi} et~al.}{2006}]{giommicmb}
{Giommi} P.,  {Colafrancesco} S.,  {Cavazzuti} E.,  {Perri} M.,   {Pittori} C.,
   2006, \mn@doi [\aap] {10.1051/0004-6361:20053402}, \href
  {http://adsabs.harvard.edu/abs/2006A%26A...445..843G} {445, 843}

\bibitem[\protect\citeauthoryear{{Giommi}, {Colafrancesco}, {Padovani},
  {Gasparrini}, {Cavazzuti}  \& {Cutini}}{{Giommi} et~al.}{2009}]{pg2009}
{Giommi} P.,  {Colafrancesco} S.,  {Padovani} P.,  {Gasparrini} D.,
  {Cavazzuti} E.,   {Cutini} S.,  2009, \mn@doi [\aap]
  {10.1051/0004-6361/20078905}, \href
  {http://adsabs.harvard.edu/abs/2009A%26A...508..107G} {508, 107}

\bibitem[\protect\citeauthoryear{{Kaiser} et~al.,}{{Kaiser}
  et~al.}{2010}]{Kaiser2010}
{Kaiser} N.,  et~al., 2010, in Society of Photo-Optical Instrumentation
  Engineers (SPIE) Conference Series. , \mn@doi{10.1117/12.859188}

\bibitem[\protect\citeauthoryear{Landt, Padovani, Perlman, Giommi, Bignall  \&
  Tzioumis}{Landt et~al.}{2001}]{L01}
Landt H.,  Padovani P.,  Perlman E.~S.,  Giommi P.,  Bignall H.,   Tzioumis A.,
   2001, MNRAS, 323, 757

\bibitem[\protect\citeauthoryear{{Lusso} \& {Risaliti}}{{Lusso} \&
  {Risaliti}}{2016}]{LR2016}
{Lusso} E.,  {Risaliti} G.,  2016, \mn@doi [\apj]
  {10.3847/0004-637X/819/2/154}, \href
  {http://adsabs.harvard.edu/abs/2016ApJ...819..154L} {819, 154}

\bibitem[\protect\citeauthoryear{{Madrid} \& {Macchetto}}{{Madrid} \&
  {Macchetto}}{2009}]{madridmacchetto}
{Madrid} J.~P.,  {Macchetto} D.,  2009, in Bulletin of the American
  Astronomical Society. pp 913--914 (\mn@eprint {arXiv} {0901.4552})

\bibitem[\protect\citeauthoryear{{Mao}, {Urry}, {Marchesini}, {Landoni},
  {Massaro}  \& {Ajello}}{{Mao} et~al.}{2017}]{addconcl4}
{Mao} P.,  {Urry} C.~M.,  {Marchesini} E.,  {Landoni} M.,  {Massaro} F.,
  {Ajello} M.,  2017, \mn@doi [\apj] {10.3847/1538-4357/aa74b8}, \href
  {http://adsabs.harvard.edu/abs/2017ApJ...842...87M} {842, 87}

\bibitem[\protect\citeauthoryear{{Merloni} et~al.,}{{Merloni}
  et~al.}{2012}]{erosita}
{Merloni} A.,  et~al., 2012, arXiv e-prints, \href
  {http://adsabs.harvard.edu/abs/2012arXiv1209.3114M} {}

\bibitem[\protect\citeauthoryear{{Mignani}}{{Mignani}}{2009}]{Mignani2009}
{Mignani} R.~P.,  2009, preprint, \href
  {http://adsabs.harvard.edu/abs/2009arXiv0902.0634M} {} (\mn@eprint {arXiv}
  {0902.0634})

\bibitem[\protect\citeauthoryear{{Monet} et~al.,}{{Monet} et~al.}{2003}]{USNOB}
{Monet} D.~G.,  et~al., 2003, \mn@doi [AJ] {10.1086/345888}, \href
  {http://adsabs.harvard.edu/cgi-bin/nph-bib_query?bibcode=2003AJ....125..984M&db_key=AST}
  {125, 984}

\bibitem[\protect\citeauthoryear{Moretti, Campana, Lazzati  \&
  Tagliaferri}{Moretti et~al.}{2003}]{Moretti03}
Moretti A.,  Campana S.,  Lazzati D.,   Tagliaferri G.,  2003, ApJ, 588, 696

\bibitem[\protect\citeauthoryear{{Moretti} et~al.,}{{Moretti}
  et~al.}{2006}]{Swifterror}
{Moretti} A.,  et~al., 2006, \mn@doi [\aap] {10.1051/0004-6361:200600007},
  \href {http://adsabs.harvard.edu/abs/2006A%26A...448L...9M} {448, L9}

\bibitem[\protect\citeauthoryear{Padovani \& Giommi}{Padovani \&
  Giommi}{1995}]{P95}
Padovani P.,  Giommi P.,  1995, ApJ, 444, 567

\bibitem[\protect\citeauthoryear{{Padovani}, {Giommi}, {Landt}  \&
  {Perlman}}{{Padovani} et~al.}{2007}]{Padovani2007}
{Padovani} P.,  {Giommi} P.,  {Landt} H.,   {Perlman} E.~S.,  2007, \mn@doi
  [\apj] {10.1086/516815}, \href
  {http://adsabs.harvard.edu/abs/2007ApJ...662..182P} {662, 182}

\bibitem[\protect\citeauthoryear{{Perley} et~al.,}{{Perley}
  et~al.}{2009}]{Perley2009}
{Perley} R.,  et~al., 2009, \mn@doi [IEEE Proceedings]
  {10.1109/JPROC.2009.2015470}, \href
  {http://adsabs.harvard.edu/abs/2009IEEEP..97.1448P} {97, 1448}

\bibitem[\protect\citeauthoryear{{Perlman}, {Padovani}, {Giommi}, {Sambruna},
  {Jones}, {Tzioumis}  \& {Reynolds}}{{Perlman} et~al.}{1998}]{deepxrayradio}
{Perlman} E.~S.,  {Padovani} P.,  {Giommi} P.,  {Sambruna} R.,  {Jones} L.~R.,
  {Tzioumis} A.,   {Reynolds} J.,  1998, \mn@doi [\aj] {10.1086/300283}, \href
  {http://adsabs.harvard.edu/abs/1998AJ....115.1253P} {115, 1253}

\bibitem[\protect\citeauthoryear{{Piranomonte}, {Perri}, {Giommi}, {Landt}  \&
  {Padovani}}{{Piranomonte} et~al.}{2007}]{PaperIII}
{Piranomonte} S.,  {Perri} M.,  {Giommi} P.,  {Landt} H.,   {Padovani} P.,
  2007, \mn@doi [\aap] {10.1051/0004-6361:20077086}, \href
  {http://adsabs.harvard.edu/abs/2007A\%26A...470..787P} {470, 787}

\bibitem[\protect\citeauthoryear{{Puccetti} et~al.,}{{Puccetti}
  et~al.}{2011}]{Swift}
{Puccetti} S.,  et~al., 2011, \mn@doi [\aap] {10.1051/0004-6361/201015560},
  \href {http://adsabs.harvard.edu/abs/2011A%26A...528A.122P} {528, A122}

\bibitem[\protect\citeauthoryear{Rector, Stocke, Perlman, Morris  \&
  Gioia}{Rector et~al.}{2000}]{Rec00}
Rector T.~A.,  Stocke J.~T.,  Perlman E.~S.,  Morris S.~L.,   Gioia I.~M.,
  2000, AJ, 120, 1626

\bibitem[\protect\citeauthoryear{{Richter}}{{Richter}}{1975}]{stat1}
{Richter} G.~A.,  1975, Astronomische Nachrichten, \href
  {http://adsabs.harvard.edu/abs/1975AN....296...65R} {296, 65}

\bibitem[\protect\citeauthoryear{Rosati et~al.,}{Rosati
  et~al.}{2002}]{Rosati02}
Rosati P.,  et~al., 2002, ApJ, 566, 667

\bibitem[\protect\citeauthoryear{{Savaglio} \& {Grothkopf}}{{Savaglio} \&
  {Grothkopf}}{2013}]{Savaglio2013}
{Savaglio} S.,  {Grothkopf} U.,  2013, \mn@doi [\pasp] {10.1086/670027}, \href
  {http://adsabs.harvard.edu/abs/2013PASP..125..287S} {125, 287}

\bibitem[\protect\citeauthoryear{{Singh}, {Barrett}, {White}, {Giommi}  \&
  {Angelini}}{{Singh} et~al.}{1995}]{wgacat}
{Singh} K.~P.,  {Barrett} P.,  {White} N.~E.,  {Giommi} P.,   {Angelini} L.,
  1995, \mn@doi [\apj] {10.1086/176595}, \href
  {http://adsabs.harvard.edu/abs/1995ApJ...455..456S} {455, 456}

\bibitem[\protect\citeauthoryear{Stickel, Fried, K\"uhr, Padovani  \&
  Urry}{Stickel et~al.}{1991}]{Sti91}
Stickel M.,  Fried J.~W.,  K\"uhr H.,  Padovani P.,   Urry C.~M.,  1991, ApJ,
  374, 431

\bibitem[\protect\citeauthoryear{{Sutherland} \& {Saunders}}{{Sutherland} \&
  {Saunders}}{1992}]{stat2}
{Sutherland} W.,  {Saunders} W.,  1992, \mnras, \href
  {http://adsabs.harvard.edu/abs/1992MNRAS.259..413S} {259, 413}

\bibitem[\protect\citeauthoryear{{Turriziani}, {Cavazzuti}  \&
  {Giommi}}{{Turriziani} et~al.}{2007}]{roxa}
{Turriziani} S.,  {Cavazzuti} E.,   {Giommi} P.,  2007, \mn@doi [\aap]
  {10.1051/0004-6361:20077114}, \href
  {http://adsabs.harvard.edu/abs/2007A%26A...472..699T} {472, 699}

\bibitem[\protect\citeauthoryear{{Tyson}, {Ivezic}, {Strauss}  \& {LSST Science
  Collaborations}}{{Tyson} et~al.}{2012}]{Ivezic2012}
{Tyson} J.~A.,  {Ivezic} Z.,  {Strauss} M.,   {LSST Science Collaborations}
  2012, in American Astronomical Society Meeting Abstracts 219. p. 156.05

\bibitem[\protect\citeauthoryear{Urry \& Padovani}{Urry \&
  Padovani}{1995}]{Urry95}
Urry C.~M.,  Padovani P.,  1995, PASP, 107, 803

\bibitem[\protect\citeauthoryear{{Vagnetti}, {Turriziani}, {Trevese}  \&
  {Antonucci}}{{Vagnetti} et~al.}{2010}]{vagn10}
{Vagnetti} F.,  {Turriziani} S.,  {Trevese} D.,   {Antonucci} M.,  2010,
  \mn@doi [\aap] {10.1051/0004-6361/201014320}, \href
  {http://adsabs.harvard.edu/abs/2010A%26A...519A..17V} {519, A17}

\bibitem[\protect\citeauthoryear{{Voges} et~al.,}{{Voges}
  et~al.}{1999}]{rosatbright}
{Voges} W.,  et~al., 1999, \aap, \href
  {http://adsabs.harvard.edu/abs/1999A%26A...349..389V} {349, 389}

\bibitem[\protect\citeauthoryear{{Voges} et~al.,}{{Voges}
  et~al.}{2000}]{rosatfaint}
{Voges} W.,  et~al., 2000, \iaucirc, \href
  {http://adsabs.harvard.edu/abs/2000IAUC.7432....3V} {7432, 3}

\bibitem[\protect\citeauthoryear{Wall \& Peacock}{Wall \&
  Peacock}{1985}]{Wall85}
Wall J.~V.,  Peacock J.~A.,  1985, MNRAS, 216, 173

\bibitem[\protect\citeauthoryear{{Wall}, {Jackson}, {Shaver}, {Hook}  \&
  {Kellermann}}{{Wall} et~al.}{2005}]{Wall04}
{Wall} J.~V.,  {Jackson} C.~A.,  {Shaver} P.~A.,  {Hook} I.~M.,   {Kellermann}
  K.~I.,  2005, \mn@doi [\aap] {10.1051/0004-6361:20041786}, \href
  {http://adsabs.harvard.edu/abs/2005A%26A...434..133W} {434, 133}

\bibitem[\protect\citeauthoryear{Wolter \& Celotti}{Wolter \&
  Celotti}{2001}]{Wol01b}
Wolter A.,  Celotti A.,  2001, A\&A, 371, 527

\bibitem[\protect\citeauthoryear{{Wong} \& {Melatos}}{{Wong} \&
  {Melatos}}{2002}]{mmATCA}
{Wong} T.,  {Melatos} A.,  2002, \mn@doi [\pasa] {10.1071/AS02015}, \href
  {http://adsabs.harvard.edu/abs/2002PASA...19..475W} {19, 475}

\bibitem[\protect\citeauthoryear{{Worswick}, {Atad-Ettedgui}, {Casali}  \&
  {Henry}}{{Worswick} et~al.}{2000}]{VISTA}
{Worswick} S.~P.,  {Atad-Ettedgui} E.,  {Casali} M.~M.,   {Henry} D.~M.,  2000,
  in {P.~Dierickx} ed.,  Society of Photo-Optical Instrumentation Engineers
  (SPIE) Conference Series Vol. 4003, Society of Photo-Optical Instrumentation
  Engineers (SPIE) Conference Series. pp 373--380

\bibitem[\protect\citeauthoryear{{{\.Z}ywucka}, {Goyal}, {Jamrozy}, {Stawarz},
  {Ostrowski}, {Koz{\l}owski}  \& {Udalski}}{{{\.Z}ywucka}
  et~al.}{2018}]{addrecent2}
{{\.Z}ywucka} N.,  {Goyal} A.,  {Jamrozy} M.,  {Stawarz} {\L}.,  {Ostrowski}
  M.,  {Koz{\l}owski} S.,   {Udalski} A.,  2018, \mn@doi [\apj]
  {10.3847/1538-4357/aae36d}, \href
  {http://adsabs.harvard.edu/abs/2018ApJ...867..131Z} {867, 131}

\bibitem[\protect\citeauthoryear{di Serego-Alighieri, Danziger, Morganti  \&
  Tadhunter}{di~Serego-Alighieri et~al.}{1994}]{diS94}
di Serego-Alighieri S.,  Danziger I.~J.,  Morganti R.,   Tadhunter C.~N.,
  1994, MNRAS, 269, 998

\makeatother
\end{thebibliography}
\input{sturriziani.bbl}


\appendix

\section{Previous Surveys for the radio LogN-LogS}\label{presur}

We give in the following some details regarding the other blazar surveys shown in Fig. \ref{rlogns} in order of decreasing radio flux limit:
\begin{enumerate}
\item The {\bf 2Jy Flat Spectrum Radio Survey} is a sample of 60 confirmed blazars \citep{diS94,Urry95} included in the 2Jy 2.7 GHz sample \citep{Wall85}, based on a complete radio flux limited survey of flat spectrum ($\alpha_r < 0.5$) sources covering the entire sky with the exclusion of the Galactic plane ($|b|>10$). The corresponding blazar space density is $0.002~deg^{-2}$ and this value is plotted as open squares in Fig. \ref{rlogns}.
\item The {\bf 1Jy ASDC-RASS-NVSS} blazar sample \citep{Giommi02} is a radio flux limited ($f_r < 1Jy$ at 1.4 GHz) sample built via a cross-correlation between the ROSAT All Sky Survey (RASS) catalog of X-ray sources \citep{rosatbright} and the subsample of NVSS sources with flux densities larger than 1 Jy. The blazars within the sample are 160 over 226 total sources. This sample was used to estimate the blazar space density above 1 Jy taking into account the RASS sky coverage, the counts were converted to 5 GHz and are plotted as open circles in Fig \ref{rlogns}.
\item The {\bf WMAP selected Blazars} sample \citep[see][]{pg2009} comprises $\sim 87\%$ of WMAP foreground source detections. The counts are show in Fig. \ref{rlogns} as filled stars and are in good agreement with other radio survey at cm wavelength, except for the point at 1 Jy, which is most likely underestimated as the WMAP catalog is incompleted at this flux limit \citep{bennett03}.
\item The {\bf Parkers 1/4Jy Flat Spectrum Sample} \citep{Wall04} is a 100\% identified radio flux limited survey at 2.7 GHz. The blazar space density inferred by this survey is 0.06 objects per square degree and is show in Fig. \ref{rlogns} as open diamonds.
\item The {\bf DXRBS (Deep X-ray Radio Blazar Survey) Sample} is a radio flux limited sample based on a double selection technique at radio and X-ray frequencies and uses optical data to refine the sample \citep[see e.g.][]{deepxrayradio,L01,Padovani2007}. The blazar space density from \cite{Padovani2007} is plotted in Fig. \ref{rlogns} as open triangles.
\item The {\bf AXN (ASDC-XMM-Newton-NVSS) Sample} \citep{giommicmb} pushes the DXRBS selection technique to fluxes down 50 mJy. The counts at 50 mJy are shown in Fig. \ref{rlogns} as a black filled circle, whereas the estimated lower limits at fainter fluxes are shown as black arrows.
\item The counts from the \textbf{Extreme HBL from the Sedentary Survey} \citep[see e.g.][]{sedentary,Gio05,PaperIII} are shown in Fig. \ref{rlogns} as filled squares. This sample is a deep ($f_r \geq 3.5~mJy$ at 1.4 GHz), 100\% identified radio flux limited sample of 150 extreme HBL objects characterized by \fxfr~ratio higher than $3 \times 10^{-10}$ \ergj. This survey does not have a direct impact on the full blazar LogN-LogS shown in Fig. \ref{rlogns} as these sources represent only a tiny fraction of the overall blazar population. However, their very high \fxfr~makes them potentially significant contributors to the Cosmic background in the high energy bands, such as X-ray, $\gamma$-ray and TeV.
\end{enumerate}


\section{Additional sources}\label{blextras}

\par Some objects were not included in the final \textit{Swift} deep GRB pointings catalog as they were just above the probability threshold fixed for detection. Despite that, we found that seven of these sources have good radio and optical counterparts.
We list them in Table \ref{extras}. Also among them there is a HSP blazar, however also in this case there is a high uncertainty in the determination of \nupeak due to lack of data.
However, we underline here that we did not use these objects in the calculation of the radio and X-ray LogN-LogS, presented in the Sections \ref{logns_radio} and \ref{logns_x}. 

\begin{table*}
\centering
\caption{{Sources above the probability threshold of the \textit{Swift} Serendipitous Survey in deep XRT GRB Fields catalog.}}
\label{extras}
\begin{tabular}{|p{1.8cm}|p{1.8cm}|p{1.4cm}|p{2.0cm}|p{1.2cm}|p{1.2cm}|p{1.0cm}|} 
\hline
RA (J2000)        & DEC (J2000)      & Flux$_{1.4GHz}$   & Flux$_{0.5-2keV}$  & Log($\nu_p)$ & Log($\nu F_{\nu}$) & LR  \\
 (deg)         & (deg)          & (mJy)             & ($10^{-14}$erg cm$^{-2}$s$^{-1}$) &      &             &      \\
\hline
02 19 33.5 & 68 44 41.7   & 4.00    & 1.02   & 15.4{$^*$}$^1$  & -11.9 & 1.165 \\
03 36 32.7 & 17 16 56.1   & 54.8    & 1.36   & 14.1{$^*$}  & -12.6 & 1.717 \\
09 30 06.7 & 16 54 31.4   & 2.10    & 0.175  & 12.6{$^*$}  & -12.6 & 1.493 \\
11 50 41.0 & 57 18 19.1   & 2.90    & 0.188  & 14.4     & -12.5 & 0.923  \\
14 36 45.9 & 27 42 31.2   & 7.50    & 0.173  & 14.8{$^*$}  & -12.2 & 1.545  \\
16 59 02.0 & 12 27 55.7   & 3.00    & 1.64   & 14.2     & -13.0 & 0.872  \\
21 55 11.3 & 16 50 58.3   & 6.40    & 1.24   & 14.8     & -12.6 & 0.040 \\ 

\hline
Notes: $^1$ HSP \\
{$^*$ Uncertain value}
\end{tabular}
\end{table*}

\bsp	
\label{lastpage}
\end{document}